\documentclass[aps,prd,10pt,onecolumn,nofootinbib,superscriptaddress,showkeys,preprintnumbers,floatfix,fleqn]{revtex4-2}

\usepackage{epsfig}
\usepackage{physics,amsmath,amssymb}
\usepackage{multirow}
\usepackage{graphicx}
\usepackage[usenames,dvipsnames,svgnames,table]{xcolor} 
\newcommand{\refsec}[1]{Sect.\,\ref{#1}}
\renewcommand{\vec}[1]{\vectorbold{#1}}

\begin{document}
\title{Effects of pair freeze-out on photon distributions in BBN epoch}

\author{Jeongyoon Choi}
\thanks{Shared first authorship: These authors contributed equally to this work.}
\affiliation{Department of Physics, Ulsan National Institute of Science and Technology (UNIST), Ulsan 44919, Republic of Korea}
\author{Dukjae Jang}
\thanks{Shared first authorship: These authors contributed equally to this work.}
\affiliation{Department of Physics and OMEG Institute, Soongsil University, Seoul 06978, Republic of Korea}
\affiliation{Department of Physics, Incheon National University, Incheon 20012, Republic of Korea}
\author{Youngshin Kwon}
\email{Corresponding author: {ykwon@hongik.ac.kr}}
\affiliation{Department of General Studies, Hongik University, Seoul 04066, Republic of Korea}
\author{Gwangeon Seong}
\affiliation{Department of Physics, Ewha Womans University, Seoul 03760, Republic of Korea}
\author{Myeong Hwan Mun}
\affiliation{Department of Physics, Kyungpook National University, Daegu 41566, Republic of Korea}
\author{Young-Min Kim}
\affiliation{Korea Astronomy and Space Science Institute, Daejeon 34055, Republic of Korea}
\author{Kyujin Kwak}
\affiliation{Department of Physics, Ulsan National Institute of Science and Technology (UNIST), Ulsan 44919, Republic of Korea}
\author{Myung-Ki Cheoun}
\affiliation{Department of Physics and OMEG Institute, Soongsil University, Seoul 06978, Republic of Korea}

\date{\today}


\begin{abstract}
We investigate the evolution of non-extensivity in the photon distribution during the Big Bang Nucleosynthesis (BBN) epoch using Tsallis statistics. Assuming a minimal deviation from the Planck distribution, we construct the perturbed Boltzmann equation for photons, including the collision terms for pair creation and annihilation processes. We analyze the possibility that these collisions could cause a slight increase in the number of high-frequency photons within the BBN era, and consequently, the primordial plasma might be temporarily placed in a state of chemical non-equilibrium. We also discuss the restoration of the photon distribution to an equilibrium state as the Universe enters the matter-dominated era. These findings, which suggest possible changes in the photon distribution during the epoch between the BBN and the recombination, offer insights that support the previously proposed ansatz solution to the primordial lithium problem in Ref.\,\cite{Jang:2018moh}.
\end{abstract}

\pacs{}
\keywords{Big bang nucleosynthesis, primordial lithium problem, non-extensive statistics}                 

\maketitle


\section{Introduction}
\label{s01}
Standard Big Bang Nucleosynthesis (SBBN) refers to the prevailing theoretical framework that describes the production of light elements in the first few minutes after the Big Bang. It provides a detailed account of the formation of light elements such as deuterium ($\mathrm{D}$), helium isotopes (${}^3\mathrm{He}$ and ${}^4\mathrm{He}$), and lithium (${}^7\mathrm{Li}$). Remarkably, the primordial abundances of deuterium and helium isotopes are well explained by SBBN. The success of SBBN in predicting the observed abundances of these light elements, along with the cosmic microwave background (CMB) observations, lends support to the hot Big Bang theory and sheds light on the conditions of the early Universe. However, the so-called primordial lithium problem, which indicates a discrepancy between the theoretical prediction and the observational data of primordial ${}^7\mathrm{Li}$ abundance, suggests that the SBBN model may need to be revised or updated. 

To address this conundrum, various avenues have been explored in prior research. These include the investigation of the potential role of long-lived exotic particles as a solution \cite{Kusakabe:2014moa} and the examination of possible observational biases that could affect the determination of the primordial lithium abundance \cite{Howk:2012rb}. Moreover, the implications of adopting Tsallis statistics have been examined, leading to discussions about changes in the nuclei distribution function \cite{Bertulani:2012sv, Hou:2017uap, Kusakabe:2018dzx}. By introducing the entropic index $q$, the sole additional parameter for Tsallis statistics, it becomes feasible to adjust the high-energy tail of the nuclei distribution functions and the related reaction rates. However, it has been found that the application of Tsallis statistics to nuclei distribution does not fully resolve the primordial lithium problem \cite{Kusakabe:2018dzx}.

Recently, some of the current authors~\cite{Jang:2018moh} proposed that taking nonideal properties of the plasma into account could be a clue to elucidate this problem of BBN. SBBN assumes that BBN occurs within a thermal plasma, and describes how the constituents of primordial plasma are distributed in terms of their velocity or energy by using Boltzmann-Gibbs (BG) statistics. Because this assumption was supported by rapid interactions between the plasma constituents, which have characteristic time-scales shorter than the cosmic expansion, the primordial plasma in an equilibrium state has won widespread acceptance. However, we took note of the electron chemical potential increasing dramatically at $T\lesssim 10^9\,\mathrm{K}$, which might be able to lead the primordial plasma to a weakly nonideal state. Then, we adopted the Tsallis statistics for describing photon distribution to reflect the nonideal effect, which appears more appropriate than the standard BG statistics. (For more details on the model construction, refer to Ref.\,\cite{Jang:2018moh}.) We found that the abundances of the light elements up to ${}^7\mathrm{Li}$ calculated with the Tsallis distribution of photons can explain the observational data within errors when a proper value of $q$ is chosen. Although this solution is effective for the primordial lithium problem, it raises a crucial question: whether or how using Tsallis statistics is physically valid for the photon distribution in the BBN era. The present work aims at finding the answer to that question by constructing and solving the Boltzmann equation for photons with relevant collision terms.

When the temperature of the Universe reached about $10^9\,\mathrm{K}$, the energy of most photons would no longer be sufficient for electron-positron pair production ($2\gamma\to e^+e^-$). It disrupted the detailed balance between the pair production and annihilation. While the pair annihilation ($e^+e^-\to 2\gamma$) continued until consuming all positrons and leaving slightly excess electrons, the pair production could not replenish them. Accordingly, the chemical equilibrium between the pairs and photons is not established, at least for a while, and we hypothesize that therein lies the last piece of the puzzle for primordial abundances. In this study, we examine if the broken detailed balance between the pair production and annihilation processes can induce a deviation from the Planck distribution for photons by adding those processes into the collision term of the Boltzmann equation. 

To achieve a comprehensive understanding of the modifications in photon distribution during the BBN epoch, it is essential to solve the Boltzmann equation by taking into account all pertinent constituents, namely photons, electrons, positrons, and nuclei. This task would be inherently complex because of the intricate nature of the system and the coupled set of partial differential equations. To mitigate this complexity, we introduce an approximation where only photons deviate from equilibrium to a small extent. In other words, all particles except photons are in equilibrium, and a change in $q$, which describes the Tsallis statistics of photons, is small.

In principle, a perturbation in the photon distribution should feed back into the electron–positron sector, and a fully consistent analysis would require evolving the fermionic distributions dynamically. Within the accuracy of our first-order treatment, however, this effect is negligible for the following two reasons. First, the photon distortion is intrinsically small, $\delta f_\gamma/f_\gamma \ll 1$. Writing the photon number density as $n_\gamma = n_\gamma^{(0)} + n_\gamma^{(1)}$ (where $n_\gamma^{(1)}/n_\gamma^{(0)} \simeq \delta f_\gamma/f_\gamma$), we see that only this small non-thermal component can drive the fermionic sector out of equilibrium, while the overwhelming majority of scatterings involve equilibrium photons that continuously restore the electrons and positrons to a Fermi–Dirac form. Second, in the temperature range relevant to our analysis, electrons and positrons are non-relativistic, so a photon can impart only a small fractional momentum change per scattering, of order $\sqrt{T/m_e}$, which further reduces the cumulative influence of the non-thermal photons on the fermionic sector. These features together are fully consistent with the expectation that, regardless of the microscopic mechanism that might generate a non-thermal photon component, the coupled $e^\pm\!-\!\gamma$ plasma naturally relaxes back toward a common thermal equilibrium.

With this assumption, we expand the generalized Planck distribution utilizing Tsallis statistics to first order, subsequently deriving the first-order Boltzmann equation. Our analysis based on the perturbed Boltzmann equation suggests that the pair annihilation process during the BBN epoch can potentially augment $q$. Furthermore, we provide a discussion for the distorted photon distribution restoring to a Planck distribution observed during the matter-dominated era.

The remainder of this paper is organized as follows. In \refsec{s02}, we present the formalism utilized to describe the evolution of the parameter $q$ through the perturbed Boltzmann equation. In \refsec{s03}, we focus on the evaluation of the collision term, particularly emphasizing key collision processes such as pair creation and annihilation.  In \refsec{s04}, we present our results regarding the impact of Tsallis statistics on the photon distribution during the BBN epoch. Finally, \refsec{s05} provides a summary and conclusion of this paper. Note that the natural unit of $\hbar=c=1$ is used throughout this paper.

\section{Boltzmann equation}
\label{s02}
The evolution of a particle distribution is generally governed by the Boltzmann equation, represented as $\mathbb{L}[f]=\mathbb{C}[f]$. The Liouville operator, $\mathbb{L}$, is associated with the streaming part of the system in the absence of collisions. It captures the system's behavior as it approaches equilibrium or deviates from it by directing the time evolution of the distribution function. On the other hand, the collision operator, $\mathbb{C}$, accounts for interactions between particles in the system. The streaming part for the evolution of photon distribution, $f^{}_\gamma$, is detailed in the relativistic Boltzmann equation as follows:
  \begin{equation}
    E^{}_\gamma \frac{\partial f^{}_\gamma}{\partial t} - H |\vec{p}^{}_\gamma|^2 \frac{\partial f^{}_\gamma}{\partial E^{}_\gamma} = \mathbb{C}[ f^{}_\gamma(\vec{p^{}_\gamma}) ]~,
  \label{eq:bol_general}
  \end{equation}
where $H$ denotes the cosmic expansion rate.

\subsection{Perturbed photon distribution}
The model involved in this study accounts for the possibility of the deviation of photon distribution, which can then be written and generalized according to the Tsallis statistics \cite{Tsallis:1987eu, Tirnakl:1997}:
  \begin{equation}
    f^{}_{\gamma}(E^{}_\gamma,T_q;q) = \left[\left( 1- (1-q)\frac{E^{}_\gamma}{T_q} \right)^{\frac{1}{q-1}}-1\right]^{-1}~,
  \label{eq:Tsallis}
  \end{equation}
as a function of energy ($E^2_\gamma=\vec{p}^2_\gamma$ for photons) and temperature. Here, we introduce a $q$-dependent effective temperature, $T_q$, to distinguish it from the equilibrium temperature, $T$. The entropic index $q$ quantifies the deviation from equilibrium. In the limit of $q\to1$, Eq.\,\eqref{eq:Tsallis} converges to the Planck distribution with $T$. It is worth noting that the solution obtained in Ref.\,\cite{Jang:2018moh} for the primordial lithium problem was $q\approx1.027$. Considering such a slight deviation of $q$ from unity, it stands to reason that the photons are assumed to be nearly in a global equilibrium. Consequently, Eq.\,\eqref{eq:Tsallis} can be expanded in terms of perturbations around the equilibrium distribution:
  \begin{equation}
    f^{}_\gamma(E^{}_\gamma,T_q;Q) \simeq f_{\gamma}^{(0)}(E^{}_\gamma,T) + f_{\gamma}^{(1)}(E^{}_\gamma,T_q;Q)~.
  \label{eq:pert}
  \end{equation}
Here, the entropic index, $Q \equiv 1-q$, is redefined for ease of use as a perturbative parameter. The first term on the right-hand side of Eq.\,\eqref{eq:pert} stands for nothing but the Planckian form:
  \begin{equation}
    f_{\gamma}^{(0)}(E^{}_\gamma,T) =  f^{}_\gamma(E^{}_\gamma,T_q;Q\to 0) = \frac{1}{e^{E^{}_\gamma/T}-1}~,
  \label{eq:dist_1}
  \end{equation}
and the second term is specified by the perturbation truncated at first order:
  \begin{align}
    f_{\gamma}^{(1)}(E^{}_\gamma,T_q;Q) = Q\left[ \frac{\partial f^{}_\gamma(E^{}_\gamma,T_q;Q)}{\partial Q} \right]_{Q\to 0} 
    &= -\frac{Q}{2 T^2} \,\frac{E^{}_\gamma\,e^{E^{}_\gamma/T}}{(e^{E^{}_\gamma/T}-1)^2}\left( E^{}_\gamma - 2\left[\frac{dT_q}{dQ}\right]_{Q\to0} \right) \nonumber \\
     &= -\frac{Q}{2 T^2} \,\frac{E^{}_\gamma\,e^{E^{}_\gamma/T}}{(e^{E^{}_\gamma/T}-1)^2}\left( E^{}_\gamma - \frac{450 \zeta(5) T}{\pi^4} \right)~,
  \label{eq:dist_2}
  \end{align}
where $\zeta(x)$ is the Riemann zeta function.

\subsection{Photon number density}
By inserting Eq.\,\eqref{eq:pert} into Eq.\,\eqref{eq:bol_general} and performing an integration over the momentum space, we can transform the Boltzmann equation into an equation governing the evolution of the number density:
  \begin{equation}
    \frac{d \left( n_{\gamma}^{(0)} + n_{\gamma}^{(1)} \right)}{dt} + 3 H \left( n_{\gamma}^{(0)} + n_{\gamma}^{(1)} \right) 
    = \frac{g_\gamma}{(2\pi)^3} \int \mathbb{C}[f^{(0)}_\gamma + f^{(1)}_\gamma] \,\frac{d \vec{p}^{}_\gamma}{E^{}_\gamma}~,
  \label{eq:bol_n}
  \end{equation}
where $g^{}_\gamma$ represents the statistical degrees of freedom for photons and is given as $g^{}_\gamma = 2$. The zeroth- and first-order number densities are written, respectively, as follows:
  \begin{align}
    n_{\gamma}^{(0)} &= \frac{2}{(2\pi)^3}\int d\vec{p}^{}_\gamma \, f_{\gamma}^{(0)}(E^{}_\gamma,T) = \frac{2\zeta(3)T^3}{\pi^2}~, \\
    n_{\gamma}^{(1)} &= \frac{2}{(2\pi)^3}\int d\vec{p}^{}_\gamma \, f_{\gamma}^{(1)}(E^{}_\gamma,T_q;Q) = -\mathcal{A}T^3Q ~,
  \label{eq:n1}
  \end{align}
where
  \begin{equation}
   \mathcal{A} = \frac{2\pi^2}{15}-\frac{1350\zeta(5)\zeta(3)}{\pi^6}~.
  \label{eq:F}
  \end{equation}

\subsection{Linear collision term}
The general form of the collision integral $\mathbb{C}[f^{}_\gamma]$ for photon-induced processes is given in Ref.\,\cite{Kolb:1990vq}:
  \begin{align}
    \frac{g^{}_\gamma}{(2\pi)^3}\int\mathbb{C}[f_\gamma]\,\frac{d\vec{p}^{}_\gamma}{E^{}_\gamma} &= \int \left( \prod_{i=\gamma, 1, 2, 3} d\Pi^{}_i \right) \left| \mathcal{M} \right|^2  \nonumber \\
	&\quad\times(2\pi)^4 \, \delta^4 (p^{}_\gamma + p^{}_1 - p^{}_2 - p^{}_3) \,F(\vec{p}^{}_\gamma, \vec{p}^{}_1; \vec{p}^{}_2, \vec{p}^{}_3)  
  \label{eq:collision}
  \end{align}
with
  \begin{equation}
    d\Pi^{}_i = \frac{g^{}_i}{(2\pi)^3}\frac{d\vec{p}^{}_i}{2E^{}_i}~.
  \end{equation}
Here $g^{}_i$ is the statistical degrees of freedom for species $i$. $F(\vec{p}^{}_\gamma, \vec{p}^{}_1; \vec{p}^{}_2, \vec{p}^{}_3)$ is defined as:
  \begin{align}
    F(\vec{p}^{}_\gamma, \vec{p}^{}_1; \vec{p}^{}_2, \vec{p}^{}_3) &\equiv f^{}_2(\vec{p}^{}_2) f^{}_3 (\vec{p}^{}_3) \Big( 1 + f^{}_\gamma(\vec{p}^{}_\gamma) \Big) \Big( 1 \pm f^{}_1(\vec{p}^{}_1) \Big) \nonumber \\
    &\quad-f^{}_\gamma(\vec{p}^{}_\gamma) f^{}_1 (\vec{p}^{}_1) \Big( 1 \pm f^{}_2(\vec{p}^{}_2) \Big) \Big( 1 \pm f^{}_3(\vec{p}^{}_3) \Big)~,
  \label{eq:Fabc}
  \end{align}
where the positive and negative signs correspond to bosons and fermions, respectively. 

By substituting Eq.\,\eqref{eq:pert} into Eq.\,\eqref{eq:Fabc}, one can find that the collision term is linear with respect to each order, that is, $\mathbb{C}[f_\gamma^{(0)} + f_\gamma^{(1)}] = \mathbb{C}[f_\gamma^{(0)}] + \mathbb{C}[f_\gamma^{(1)}]$. The linearity of the collision term significantly simplifies the problem, enabling us to consider the effects of each order of the distribution function independently. Therefore, under the assumption that the zeroth order remains in equilibrium, such that $\mathbb{L}[f_\gamma^{(0)}] = \mathbb{C}[f_\gamma^{(0)}]$, Eq.\,\eqref{eq:bol_n} can be rewritten as follows:
  \begin{equation}
    \frac{dn_{\gamma}^{(1)}}{dt} + 3 Hn_{\gamma}^{(1)} = \frac{g_\gamma}{(2\pi)^3} \int \mathbb{C}[f^{(1)}_\gamma (\vec{p}^{}_\gamma)]\, \frac{d \vec{p}^{}_\gamma}{E^{}_\gamma}~.
  \label{eq:bol_n2}
  \end{equation}

\subsection{Liouville term}
The evolution of $n_\gamma^{(1)}$ in Eq.\,\eqref{eq:bol_n2} can be converted into a differential equation for $Q$. This subsection aims to reformulate the Liouville term in Eq.\,\eqref{eq:bol_n2} into an expression involving $Q$. To achieve this, it is essential to establish the time-temperature relation, which can be derived from the Friedmann equation. Owing to energy conservation governing the photon energy density, 
$H$ is expressed in its standard form:
  \begin{equation}
    H^2 = \left( \frac{\dot{a}}{a} \right)^2 = \frac{8\pi G}{3}\rho^{}_\text{total}~,
  \label{eq:H}
  \end{equation}
where $a$ represents the scale factor, and $\dot{a}$ is its time derivative. By using the relation between the energy density and temperature in the radiation-dominated era, it can be rewritten by
  \begin{equation}
    H = \frac{T^2}{m_\text{pl}}\sqrt{\frac{8\pi^3 g^{}_\ast}{90}}\equiv \mathcal{B}T^2~,
  \label{eq:H_re}
  \end{equation}
where $g^{}_\ast$ represents the total degrees of freedom for the effectively relativistic particles and $m_\text{pl}$ stands for the Planck mass. Since the scale factor follows $a \propto t^{1/2}$ in the radiation-dominated era, the time-temperature relation is simply as follows:
  \begin{equation}
    t = \frac{1}{2H} =\frac{1}{2\mathcal{B}T^2}~,
  \label{eq:t}
  \end{equation}
which gives rise to the time-derivative of $T$:
  \begin{equation}
    \frac{dT}{dt} = -\mathcal{B}T^3~.
  \label{eq:dTdt}
  \end{equation}

By using the time-temperature relations of Eqs.\,\eqref{eq:t} and \eqref{eq:dTdt}, the Liouville term in Eq.\,\eqref{eq:bol_n2} takes the form of
  \begin{align}
    \frac{dn^{(1)}_\gamma}{dt} + 3H n_{\gamma}^{(1)} &= \frac{\partial n_\gamma^{(1)}}{\partial T}\frac{dT}{dt} + 3H n_{\gamma}^{(1)} \nonumber \\
    &= \left( -3\mathcal{A}T^2Q-\mathcal{A}T^3\frac{dQ}{dT} \right)\left( -\mathcal{B}T^3 \right) +3\left(\mathcal{B}T^2\right)\left(-\mathcal{A}T^3Q\right) \nonumber \\
    &= \mathcal{A}\mathcal{B}T^6\frac{dQ}{dT}~.
  \label{eq:dndt}
  \end{align}

\section{Collision term in focus}
\label{s03}
In this section, we turn our attention to the evaluation of the collision term, focusing primarily on key collision processes such as Compton scattering and pair creation and annihilation. Higher-order processes, such as double Compton scattering and bremsstrahlung, can be safely neglected, given the significant reduction in electron number density at lower temperatures.

To apply Eq.\,\eqref{eq:collision} for pair creation and annihilation, that is, $\gamma(E^{}_\gamma,\vec{p}^{}_\gamma)+\gamma(E^{}_1,\vec{p}^{}_1) \rightleftharpoons e^-(E^{}_2,\vec{p}^{}_2) + e^+(E^{}_3,\vec{p}^{}_3)$, the transition amplitude is calculated as follows:
  \begin{equation}
    |\mathcal{M} |^2_\text{pair} =  2 e^4 \left[ \frac{p^{}_\gamma \cdot p^{}_3+2m_e^2}{p^{}_\gamma \cdot p^{}_2} + \frac{p^{}_\gamma \cdot p^{}_2+2m_e^2}{p^{}_\gamma \cdot p^{}_3}
     - m_e^4 \left( \frac{1}{p^{}_\gamma \cdot p^{}_2} + \frac{1}{p^{}_\gamma \cdot p^{}_3} \right)^2 \right]~,
  \label{eq:Matrix}
  \end{equation}
and 
  \begin{align}
    F_\text{pair}(\vec{p}^{}_\gamma,\vec{p}^{}_1; \vec{p}^{}_2, \vec{p}^{}_3) &= f^{}_{e^-}(\vec{p}^{}_2) f_{e^+} (\vec{p}^{}_3) \Big( 1 + f^{}_\gamma(\vec{p}^{}_\gamma) \Big) \Big( 1 + f^{}_\gamma(\vec{p}^{}_1) \Big) \nonumber \\
    &\quad-f^{}_\gamma(\vec{p}^{}_\gamma) f^{}_\gamma(\vec{p}^{}_1) \Big( 1 - f^{}_{e^-}(\vec{p}^{}_2) \Big) \Big( 1 - f^{}_{e^+}(\vec{p}^{}_3) \Big)~.
  \label{eq:Fdist}
  \end{align}
By setting electrons and positrons to remain in equilibrium — a reasonable assumption given the small momentum transfer, as supported by Ref.\,\cite{McDermott:2018uqm} — we can rewrite Eq.\,\eqref{eq:Fdist} based on Eq.\,\eqref{eq:pert} as follows:
  \begin{equation}
    F_\text{pair}(\vec{p}^{}_\gamma,\vec{p}^{}_1; \vec{p}^{}_2, \vec{p}^{}_3) = F^{(0)}_\text{pair}(\vec{p}^{}_\gamma,\vec{p}^{}_1; \vec{p}^{}_2, \vec{p}^{}_3) 
    + F^{(1)}_\text{pair}(\vec{p}^{}_\gamma,\vec{p}^{}_1; \vec{p}^{}_2, \vec{p}^{}_3) + \mathcal{O}(Q^2)~,
  \label{eq:ff}
  \end{equation}
where each term is expressed as:
  \begin{align}
    F^{(0)}_\text{pair}(\vec{p}^{}_\gamma,\vec{p}^{}_1; \vec{p}^{}_2, \vec{p}^{}_3) &= f_{e^-}(\vec{p}^{}_2) f_{e^+} (\vec{p}^{}_3) \Big(1+f_\gamma^{(0)} (\vec{p}^{}_\gamma) \Big)\Big( 1+f_\gamma^{(0)} (\vec{p}^{}_1) \Big) \nonumber \\
    &\quad -f_\gamma^{(0)} (\vec{p}^{}_\gamma) f_\gamma^{(0)} (\vec{p}^{}_1)  \Big( 1 - f_{e^-}(\vec{p}^{}_2) \Big) \Big( 1- f_{e^+} (\vec{p}^{}_3) \Big)
  \label{eq:f0}
  \end{align}
and 
  \begin{multline}
    F^{(1)}_\text{pair}(\vec{p}^{}_\gamma,\vec{p}^{}_1; \vec{p}^{}_2, \vec{p}^{}_3) 
    = f^{}_{e^-}(\vec{p}^{}_2) f^{}_{e^+} (\vec{p}^{}_3) \left[ \left( 1 + f_\gamma^{(0)}(\vec{p}^{}_\gamma) \right) f_\gamma^{(1)}(\vec{p}^{}_1) + f_\gamma^{(1)}(\vec{p}^{}_\gamma) \left( 1 + f_\gamma^{(0)}(\vec{p}^{}_1) \right) \right] \\
     -\left[ f^{(0)}_\gamma(\vec{p}) f^{(1)}_\gamma(\vec{p}^{}_1) + f^{(1)}_\gamma(\vec{p}^{}_\gamma) f^{(0)}_\gamma(\vec{p}^{}_1) \right]  \left( 1 - f_{e^-}( \vec{p}^{}_2 ) \right) \left( 1 - f_{e^+}( \vec{p}^{}_3 ) \right) ~. 
  \label{eq:f1}
  \end{multline}

Similarly, the transition amplitude for the Compton scattering, $\gamma(E^{}_\gamma,\vec{p}^{}_\gamma)+e^-(E^{}_1,\vec{p}^{}_1) \rightleftharpoons \gamma(E^{}_2,\vec{p}^{}_2) + e^-(E^{}_3,\vec{p}^{}_3)$, can be written as:
  \begin{equation}
     |\mathcal{M} |^2_\text{cs} =  2 e^4 \left[ \frac{p^{}_1 \cdot p^{}_2+2m_e^2}{p^{}_1 \cdot p^{}_\gamma} + \frac{p^{}_1 \cdot p^{}_\gamma-2m_e^2}{p^{}_1 \cdot p^{}_2}
     + m_e^4 \left( \frac{1}{p^{}_1 \cdot p^{}_\gamma} - \frac{1}{p^{}_1 \cdot p^{}_2} \right)^2 \right]~.
  \label{eq:Matrix_comptonscattering}
  \end{equation}
The zeroth- and first-order terms of the distribution part are expressed as:
  \begin{align}
    F^{(0)}_\text{cs}(\vec{p}^{}_\gamma,\vec{p}^{}_1; \vec{p}^{}_2, \vec{p}^{}_3) &= f^{(0)}_\gamma(\vec{p}^{}_2) f_{e^-} (\vec{p}^{}_3) \Big(1+f_\gamma^{(0)} (\vec{p}^{}_\gamma) \Big)\Big( 1-f_{e^-} (\vec{p}^{}_1) \Big)  \nonumber \\
	&\quad -f_\gamma^{(0)} (\vec{p}^{}_\gamma) f_{e^-} (\vec{p}^{}_1)  \Big( 1 + f_\gamma^{(0)}(\vec{p}^{}_2) \Big) \Big( 1- f_{e^-} (\vec{p}^{}_3) \Big) 
  \label{eq:fcs0}
  \end{align}
and
  \begin{multline}
    F^{(1)}_\text{cs}(\vec{p}^{}_\gamma,\vec{p}^{}_1; \vec{p}^{}_2, \vec{p}^{}_3) = \left[f^{(1)}_\gamma(\vec{p}^{}_2) \Big(1+f^{(0)}_\gamma(\vec{p}_\gamma)\Big) +f^{(0)}_\gamma(\vec{p}_2) f^{(1)}(\vec{p}_\gamma)\right] f^{}_{e^-} (\vec{p}^{}_3) \Big(1-f_{e^-}(\vec{p}^{}_1)\Big)  \\
	\quad -\left[f^{(1)}_\gamma(\vec{p}_\gamma)\Big(1+f^{(0)}_\gamma(\vec{p}_2)\Big)+f^{(0)}_\gamma(\vec{p}_\gamma)f^{(1)}_\gamma(\vec{p}_2)\right] f_{e^-}(\vec{p}^{}_1) \Big(1-f_{e^-}(\vec{p}^{}_3)\Big) ~.  
  \label{eq:fcs1}
  \end{multline}

\begin{figure}[t]
  \centering
    \includegraphics[width=9cm]{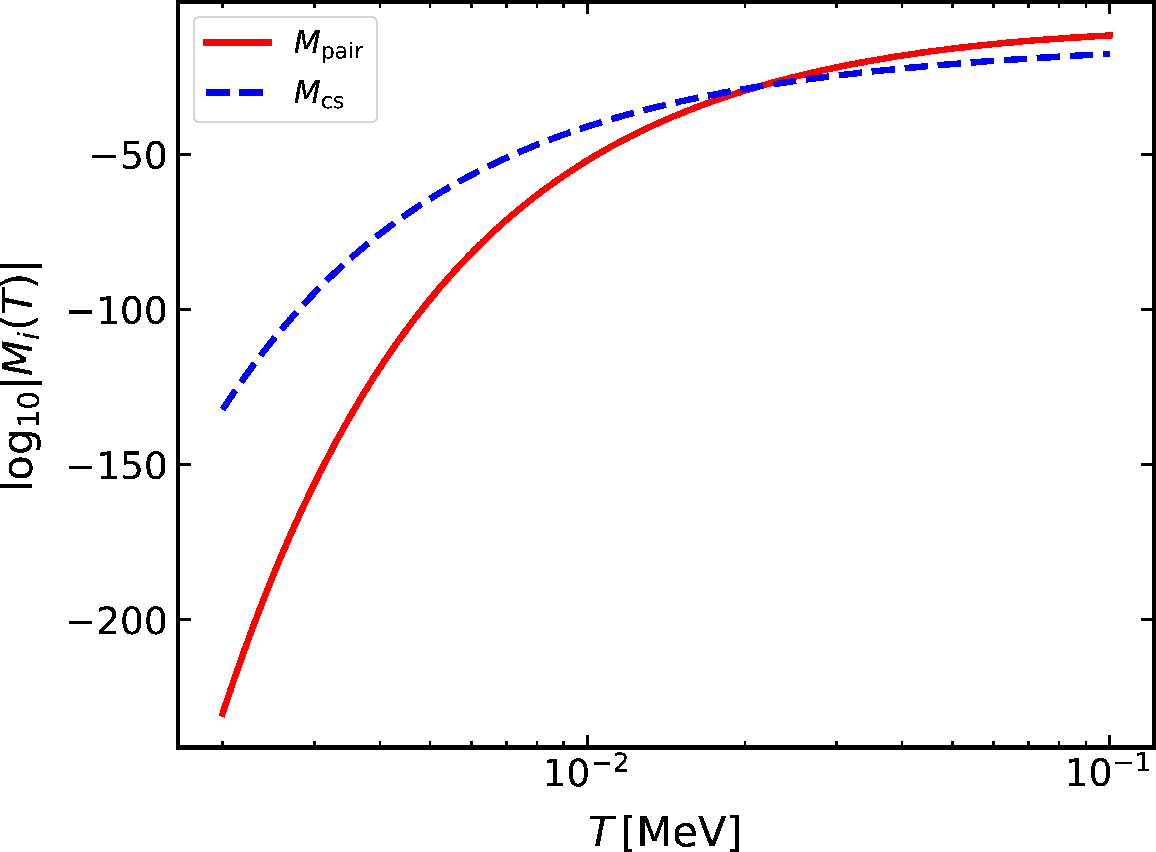}
    \caption{Logarithmic plot of the collision term $M(T)$ as a function of temperature, with $M(T)$ evaluated in $\mathrm{MeV}^4$. Note that absolute values are plotted to compare magnitudes, although the terms have opposite signs: $M_\text{pair}(T)>0$ and $M_\text{cs}(T)<0$. }
  \label{fig1}
  \end{figure}

Given that $Q$ does not depend on momentum integration, we can formulate the collision integral for the first-order perturbation as follows:
  \begin{equation}
    \frac{g^{}_\gamma}{(2\pi)^3} \int \mathbb{C}[f^{(1)}_\gamma(\vec{p}^{}_\gamma)] \frac{d \vec{p}^{}_\gamma}{p^{}_\gamma} = Q M(T)~.
  \label{eq:col_an}
  \end{equation}
Here, the right-hand side of Eq.\,\eqref{eq:col_an} stands for the following expression:
  \begin{align}
    Q M(T) &\equiv Q\left( M_\text{pair}(T) + M_\text{cs}(T) \right) \nonumber \\
    &= \int \left( \prod_{i=\gamma, 1,2,3} d\Pi_i \right) (2\pi)^4 \delta^{3} (\vec{p}^{}_\gamma + \vec{p}^{}_1 - \vec{p}^{}_2 - \vec{p}^{}_3)\, \delta (E^{}_\gamma + E^{}_1 - E^{}_2 - E^{}_3) \nonumber \\
    &\quad\times \bigg[ | \mathcal{M} |^2_\text{pair}F^{(1)}_\text{pair}(\vec{p}^{}_\gamma, \vec{p}^{}_1, \vec{p}^{}_2, \vec{p}^{}_3) +| \mathcal{M} |^2_\text{cs}F^{(1)}_\text{cs}(\vec{p}^{}_\gamma, \vec{p}^{}_1, \vec{p}^{}_2, \vec{p}^{}_3)\bigg]~.  
  \label{eq:M}
  \end{align}
The evaluation of this collision integral employs Gauss-Legendre and Gauss-Laguerre quadrature methods, following the approach described in Ref.\,\cite{Thomas:2019ran}. The magnitudes of $M_\text{pair}$ and $M_\text{cs}$ are plotted in Fig.\,\ref{fig1}. Because they have opposite signs, the total $M(T)$ changes sign from positive to negative as $T$ decreases past around $0.0212\,\mathrm{MeV}$.

By applying Eqs.\,\eqref{eq:dndt} and \eqref{eq:col_an}, we rewrite the perturbed Boltzmann equation given in Eq.\,\eqref{eq:bol_n2} as shown in below:
  \begin{equation}
    \frac{dQ}{dT} = \frac{M(T)}{\mathcal{A}\mathcal{B}T^6}Q~.
  \label{eq:dQdT}
  \end{equation}

Here we emphasize the methodological advantage of employing Tsallis statistics. In principle, the evolution of the photon distribution function, $f_\gamma(E, T)$, should be solved directly from the Boltzmann equation, even at the first-order perturbative level. However, once the Tsallis form of the distribution is adopted, the perturbed Boltzmann equation can be reduced to Eq.\,\eqref{eq:dQdT}, which is an ordinary differential equation (ODE) for the non-extensivity parameter $Q$, rather than a partial differential equation (PDE) for $f_\gamma(E, T)$. This formulation significantly simplifies the computation: by solving a single ODE for $Q(T)$, one can effectively trace the dynamical evolution of the photon distribution with much lower computational cost while retaining the essential non-equilibrium physics.

\section{Results}	
\label{s04}
\subsection{Trivial solution}
The solution of Eq.\,\eqref{eq:dQdT} is given as
  \begin{equation}
    Q(T) = Q^{}_0\exp\left[ \int_{T^{}_0}^T \frac{M(T^\prime)}{\mathcal{A}\mathcal{B}{T^\prime}^6} dT^\prime \right]~,
  \label{eq:sol}
  \end{equation}
where $Q^{}_0$ is a constant to be determined by an initial condition $Q(T^{}_0)$. Setting $Q^{}_0=0$ at a specific $T^{}_0$ leads to a trivial solution $Q=0$, which drives the system towards perfect equilibrium by rapid collisions in the BBN epoch. This trivial solution is equivalent to the assumption of SBBN. 

  \begin{figure}[t]
  \centering
    \includegraphics[width=9cm]{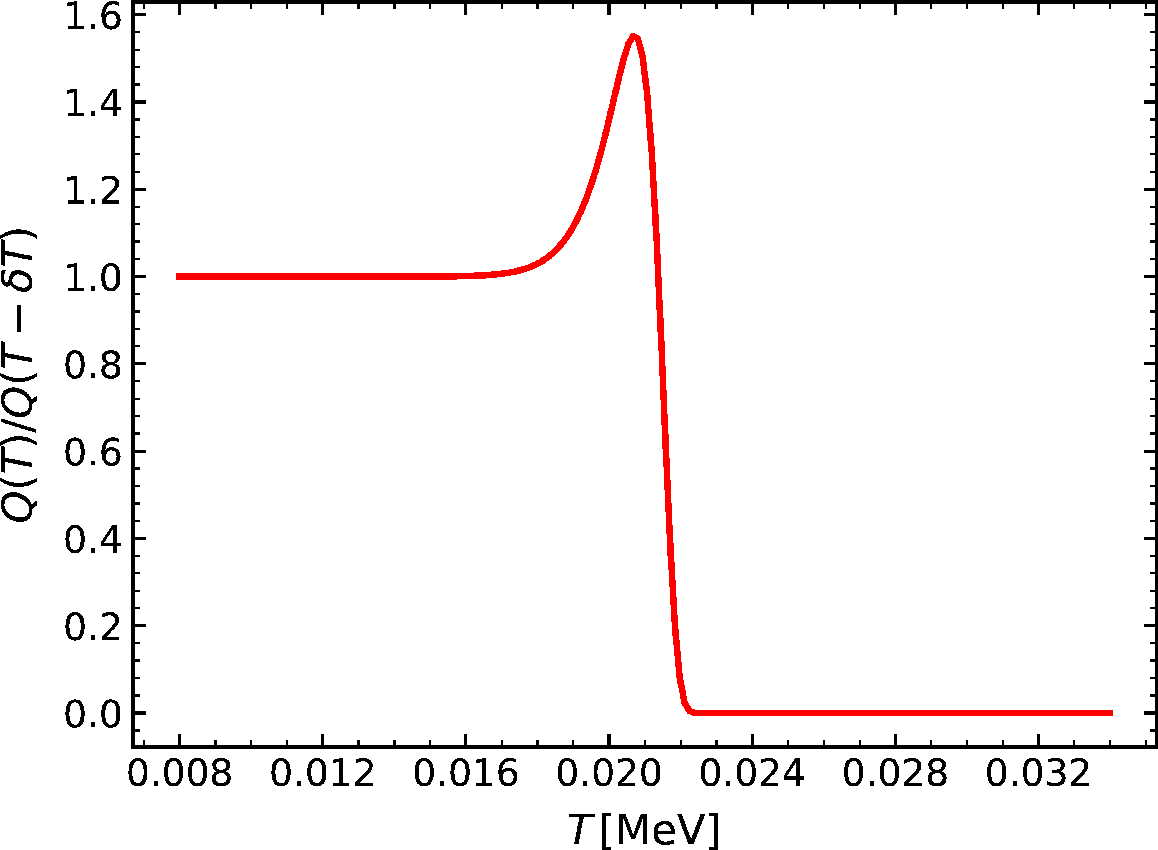}
    \caption{Ratio of $Q(T)$ for a small temperature drop. Since this ratio is independent of the initial condition, $Q^{}_0=Q(T^{}_0)$, we can assess the sensitivity of the solution to changes in temperature while assuming a non-zero $Q^{}_0$.}
  \label{fig2}
  \end{figure}

\subsection{Non-zero initial condition}
However, a perfect thermal equilibrium in the early Universe is unlikely. Even negligibly small, non-zero fluctuations could emerge because of various instabilities and other unknown factors. Minor seed fluctuations in temperature could significantly impact the evolution of the Universe. In this context, we investigate the implications of a non-zero initial condition $Q^{}_0$ in Eq.\,\eqref{eq:sol}. 

To determine $T^{}_0$ before finding $Q^{}_0$, we tracked the change in $Q(T)$ over a small decrease in temperature $\delta T$:
  \begin{equation}
    \frac{Q(T)}{Q(T-\delta T)} = \exp\left[-\int_{T}^{T-\delta T} \frac{M(T^\prime)}{\mathcal{A}\mathcal{B}{T^\prime}^6} dT^\prime\right]~.
  \label{eq:ratio}
  \end{equation}
As shown in Fig.\,\ref{fig2}, the ratio in Eq.\,\eqref{eq:ratio} is close to zero at $T\gtrsim 0.022\,\mathrm{MeV}$, suggesting that $Q(T)$ changes considerably in magnitude over small temperature shifts. Due to this strong variation, the assumption of a minor, perturbative deviation in $Q(T)$ becomes invalid. Consequently, this high temperature region cannot be chosen as the initial temperature $T^{}_0$ for the system under consideration, as such significant changes contradict the perturbative behavior of $Q$. Instead, this may be interpreted as $Q$ itself being almost zero at $T\gtrsim 0.022\,\mathrm{MeV}$. The ratio exhibits steep changes as $T$ decreases past approximately $0.022\,\mathrm{MeV}$. At $T\sim0.021\,\mathrm{MeV}$, $Q$ attains its maximum deviation and then begins to return to unity. In the lower temperature region, as the collision term approaches zero, the ratio eventually stabilizes at unity, meaning that $Q$ no longer deviates. According to this analysis, the behavior of $Q$ after $T\sim0.022\,\mathrm{MeV}$ conforms with our physical expectations. Therefore, we take the initial temperature $T^{}_0\simeq 0.022\,\mathrm{MeV}$ in Eq.\,\eqref{eq:sol}. 

\begin{figure}[t]    
  \centering
    \includegraphics[width=9cm]{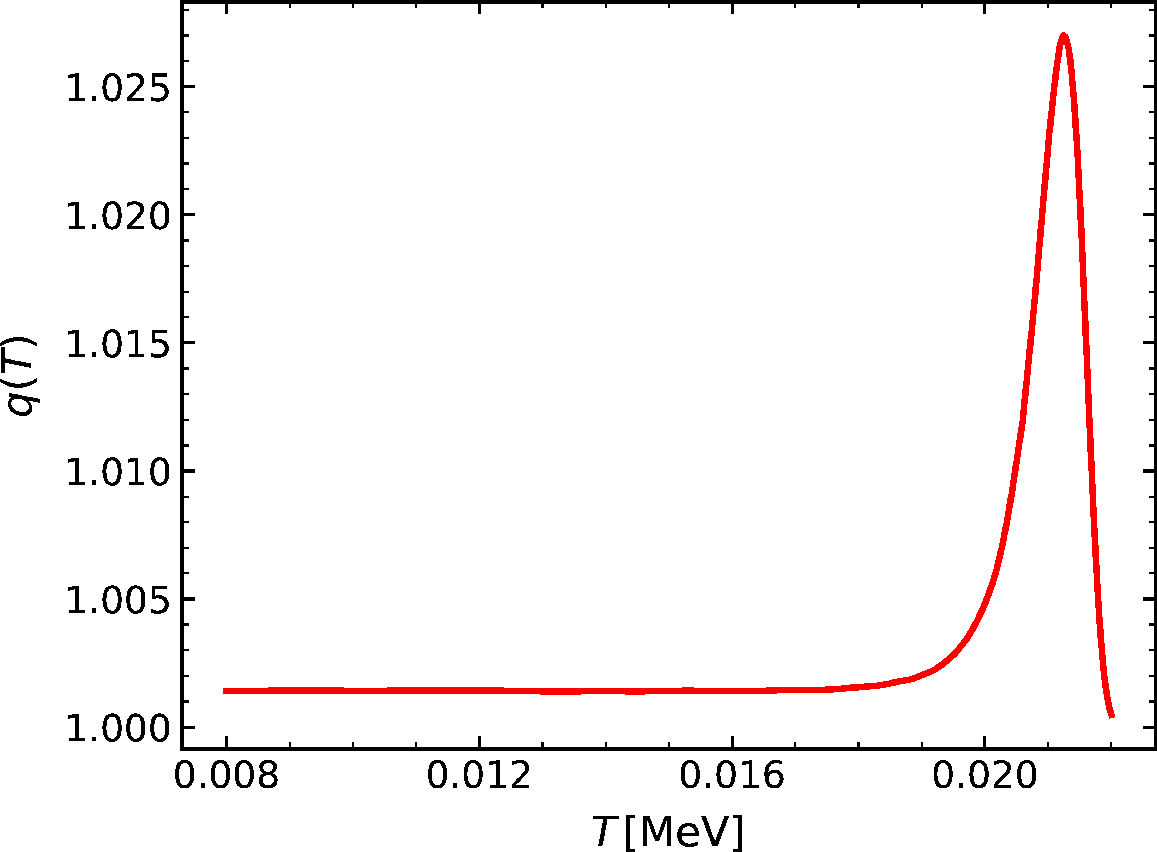}
    \caption{Solution of $q(T) = 1-Q(T)$ as a function of temperature with the initial condition of a deviation $Q^{}_0=-4.78\times10^{-4}$ at $T^{}_0=0.022\,\mathrm{MeV}$.}
  \label{fig3}
  \end{figure}

Having established $T^{}_0$, the next step is to determine the value of the initial condition $Q^{}_0$. Since we have no direct method to estimate $Q^{}_0$, we choose to determine it by requiring that the original entropic index $q$ reaches a maximum value of $q\simeq1.027$, consistent with the solution reported in Ref.\,\cite{Jang:2018moh}. Fig.\,\ref{fig3} presents, as an example, the numerical solution of $q(T)=1-Q(T)$ obtained with an initial condition of $Q^{}_0 = -4.78\times10^{-4}$ at $T^{}_0 = 0.022\,\mathrm{MeV}$. By selecting this initial condition, we ensure that our model remains consistent with empirical observations of BBN abundances, as the deviation of $q\simeq1.027$ from unity leads to a slight enhancement of the high-energy tail of the photon distribution. As indicated in Fig.\,\ref{fig3}, the initial small deviation in $q(T)$ evolves rapidly as temperature decreases. However, this increasing trend in $q(T)$ does not persist because the Compton scattering dominates the collision term $M(T)$ in magnitude, as depicted in Fig.\,\ref{fig1}. Consequently, the photon distribution equilibrates until the Universe reaches a collisionless regime at low temperatures. Thus, in Fig.\,\ref{fig3}, $q(T)$ eventually saturates back at $q\simeq 1.001$ below $T\sim 0.018\,\mathrm{MeV}$. This also explains the CMB observations, where the photon distribution closely approximates the Planck distribution \cite{Planck:2019evm}.

The freeze-out of electron-positron pairs is attributed to the disruption of the detailed balance between pair creation and annihilation. This process already begins during the early stages of BBN when the thermal energy per particle becomes insufficient to create such pairs. The small deviation in $Q^{}_0$ might represent a vestige of these early-stage BBN conditions. The pair freeze-out process could cause this small fluctuation to evolve and become pronounced enough to contribute to BBN abundances, particularly affecting the primordial lithium abundance. This would be one possible interpretation of our calculation in this work. 

\subsection{Remarks on prior study}
As mentioned earlier, Ref.\,\cite{Jang:2018moh} addressed the primordial lithium problem by using Tsallis statistics to parameterize a distorted photon distribution during the BBN epoch. It led to a specific value of $q\simeq1.027$ at $T\simeq 4\times 10^8\,\mathrm{K}$, indicating a deviation from the Planck distribution. Such a deviation was a key to resolving the primordial lithium problem. Our current work not only verifies these findings but also significantly extends them. While the resulting $q(T)$ in this study is markedly dependent on the chosen initial conditions, certain choices yield results that align with our previous study. Importantly, our studies challenge the SBBN model by suggesting that the photon distribution can deviate from the Planck distribution. It could be a crucial point that refines the SBBN and opens up new avenues for understanding the early Universe and the primordial abundances.

While the choice of initial conditions continues to be a subject for further exploration, the consistency in the qualitative behavior across both studies demonstrates the idea that deviations from the Planck distribution of photons could be intrinsic features of the early Universe. 

\section{Conclusion}
\label{s05}
In this study, we investigated the temporal evolution of the entropic index $Q$ in Tsallis statistics, employing the first-order Boltzmann equation with pertinent collision terms. Our findings indicate that during the BBN epoch, collisional processes for the freeze-out of $e^+e^-$ pairs can yield a decrement in $Q$, thereby causing a distortion in the high-energy tail of the photon distribution. To achieve a physically meaningful solution for $Q$ that resolves the primordial lithium problem, we optimized the initial condition of the Boltzmann equation, letting the initial fluctuation be small but non-vanishing. As a consequence, the decline in $Q$ ceases around $T\le0.02\,\mathrm{MeV}$, and $Q$ returns to zero because of the Compton process in the collision term, culminating in the stabilization of $Q$ during the terminal stage of the BBN epoch.

Our results imply that the variations in $Q$ predicted by our model could be detectable exclusively through BBN observations. A more comprehensive analysis, encompassing the evolution of $Q$ throughout the early cosmic history — including both the BBN epoch and subsequent eras — is essential for a deeper understanding of the photon distribution and the fundamental physics governing the early Universe.

It is noteworthy that observational evidence for small fluctuations during the BBN era is not directly available, unlike the CMB, which offers a clear snapshot of temperature fluctuations at the time of recombination. The temperature fluctuations during BBN may have been precursors to the density and temperature fluctuations observable in the CMB. Exploring the link between these two epochs can yield valuable insights into the evolution of the early Universe, particularly regarding inflationary theory and the development of cosmic structures.


\begin{acknowledgements}
D.J. was partly supported by National Natural Science Foundation of China (No. 12335009), and the National Key R$\&$D Program of China (2022YFA1602401).
M.H.M. was supported by the National Research Foundation of Korea (NRF) grant funded by the Korea government (Grant No. NRF- 2021R1F1A1060066).
M.K.C. was supported by NRF under grant numbers, NRF-2021NR060129 and NRF-202516071941.
M.H.M. and M.K.C. were also supported by NRF-2020R1A2C3006177.
K.K. was supported by NRF-2022R1F1A1073890 and 2023 UBSI Innovation Project of UNIST.
Y.M.K. was supported by NRF-2022R1C1C2012226.
\end{acknowledgements}


\end{document}